\documentclass[aps,prx,10pt,twocolumn,superscriptaddress,floatfix]{revtex4-2}

\usepackage[utf8]{inputenc}
\usepackage[T1]{fontenc}
\usepackage{bm}
\usepackage[english]{babel}	
\usepackage{breakcites}
\usepackage{lmodern}
\usepackage[normalem]{ulem}

\usepackage{ae,aecompl}
\usepackage{subfigure}
\usepackage{latexsym,stmaryrd}
\usepackage{graphicx}
\usepackage[labelfont=bf,font={small},labelsep=period]{caption}
\usepackage[hidelinks,linkcolor=blue,colorlinks=true,urlcolor=blue,citecolor=blue]{hyperref}
\usepackage{braket}
\usepackage{amsmath}
\usepackage{upgreek}
\usepackage{cleveref}
\usepackage{comment}
\usepackage{url}
\usepackage{amssymb}

\newcommand{\micro}{$\upmu$}
\newcommand{\gc}{\gamma_\mathrm{c}}
\newcommand{\ntr}{n_\mathrm{tr}}
\newcommand{\rinj}{r^\mathrm{inj}}
\newcommand{\Rin}{R^\mathrm{inj}}
\newcommand{\tnr}{\tau_\mathrm{nr}}
\newcommand{\trad}{\tau_\mathrm{rad}}
\newcommand{\tp}{\tau_\mathrm{ph}}
\newcommand{\ah}{\alpha_\mathrm{h}}
\newcommand{\coeff}{\mathcal{C}_\mathrm{abs}}
\newcommand{\nsat}{n_\mathrm{sat}}
\newcommand{\teff}{\tau_\mathrm{eff}}
\newcommand{\frep}{f_\mathrm{rep}}

\begin{document}

\title{Dynamical control of non-Hermitian coupling between sub-threshold nanolasers enables Q-switched pulse generation}

\author{Kristian Seegert}
\affiliation{Department of Electrical and Photonics Engineering, Technical University of Denmark, Ørsteds Plads 345A, 2800 Kgs. Lyngby, Denmark}
\affiliation{NanoPhoton - Center for Nanophotonics, Ørsteds Plads 345A, 2800 Kgs. Lyngby, Denmark}

\author{Roberto Gajardo}
\affiliation{Institut de Physique de Nice, CNRS, Université Côte d'Azur, 17 rue Julien Lauprêtre, 06000 Nice, France}

\author{Guillaume Huyet}
\affiliation{Institut de Physique de Nice, CNRS, Université Côte d'Azur, 17 rue Julien Lauprêtre, 06000 Nice, France}

\author{Fabrice Raineri}
\affiliation{Institut de Physique de Nice, CNRS, Université Côte d'Azur, 17 rue Julien Lauprêtre, 06000 Nice, France}

\author{Guilhem Madiot}
\affiliation{Institut de Physique de Nice, CNRS, Université Côte d'Azur, 17 rue Julien Lauprêtre, 06000 Nice, France}
\affiliation{Corresponding author: guilhem.madiot@univ-cotedazur.fr}

\begin{abstract}
Non-Hermitian photonics provides a framework to engineer the gain and loss of optical modes in open systems, enabling control of their spectral and dynamical properties. In particular, the ability to dynamically tune modal losses offers a route to implement functionalities traditionally relying on cavity Q-factor modulation, such as Q-switching, within nanophotonic platforms. Here, we demonstrate the generation of short optical pulses in a pair of phase-coupled photonic crystal nanolasers exploiting non-Hermitian coupling. Two waveguide-coupled nanocavities are operated below their individual lasing thresholds and subjected to asymmetric optical pumping, such that a transient carrier-induced detuning modifies the interference conditions between them. This dynamically controls the gain and loss of the collective modes and, upon crossing a resonance condition, leads to the rapid release of stored carrier energy as an optical pulse. A rate-equation model captures the interplay between carrier dynamics and modal coupling and reproduces the observed behavior. Experiments performed on an indium phosphide platform show pulse generation from cavities that do not lase efficiently on their own in continuous-wave operation, with temporal characteristics governed by carrier dynamics. These results indicate that non-Hermitian coupling can be used to control the effective cavity losses in time, providing a route to pulse generation in integrated photonic systems.
\end{abstract}

\maketitle

\section{Introduction}

Recent advances in non-Hermitian photonics have opened new opportunities for engineering light–matter interactions in coupled-resonator systems~\cite{dembowskiExperimentalObservationTopological2001,ozdemirParityTimeSymmetry2019,liExceptionalPointsNonHermitian2023}. 
In open systems, the interplay of gain, loss, and coupling gives rise to collective modes whose properties depend sensitively on interference conditions. In this framework, controlling modal losses becomes a central degree of freedom, and non-Hermitian coupling provides a route to dynamically tune the effective Q-factor of collective modes through detuning and interference.
This capability is directly relevant to Q-switching, which relies on the controlled modulation of cavity losses to store and release energy in the form of short optical pulses.

The ability to generate short optical pulses on demand is essential for applications ranging from high-speed optical communication~\cite{huChipbasedOpticalFrequency2021} to precision metrology~\cite{udemOpticalFrequencyMetrology2002} and neuromorphic photonic computing~\cite{shastriPhotonicsArtificialIntelligence2021}. Among the various approaches to pulse generation, Q-switching is attractive for its ability to store energy in a gain medium and release it as a short, high-peak-power burst of light. Traditionally, Q-switching has been implemented in bulk lasers through active or passive modulation of intracavity losses~\cite{mcclungGiantOpticalPulsations1962,sveltoPrinciplesLasers2010}. Translating this functionality to integrated nanophotonic platforms remains challenging due to the inherently low photon-storage capacity of nanocavities and the need for ultrafast, low-energy control mechanisms.

Several strategies have been explored in nanolaser systems, including mode-field switching in coupled cavities and cavity dumping in microscopic Fano lasers~\cite{pellegrinoModefieldSwitchingNanolasers2020,dongCavityDumpingUsing2023,morkPhotonicCrystalFano2014,morkNanostructuredSemiconductorLasers2025,morkSemiconductorFanoLasers2019}. These approaches likewise involve dynamical control of QNMs, but differ in modulation scheme and tuning mechanism, often relying on thermal or nonlinear effects. Distinct pulse-generation regimes have also been demonstrated through passive Q-switching or self-pulsing in nanophotonic lasers, relying on saturable absorption or intrinsic nonlinear dynamics~\cite{yuDemonstrationSelfpulsingPhotonic2017,delmulleSelfPulsingNanobeamPhotonic2022,liSelfpulsingDualmodeLasing2023,seegertSelfpulsingDynamicsMicroscopic2024}. More broadly, these works indicate that pulsed emission can be engineered by dynamically controlling the quasinormal modes of open photonic systems.

A key challenge in exploiting mode coupling to engineer optical functionalities is the need to independently control both frequency detuning and losses. The most common approach relies on direct evanescent coupling, mediated by spatial overlap of the cavity fields~\cite{adamsHighfrequencyDynamicsEvanescentlycoupled2019,zhangParityTimeSymmetryBreaking2016,haddadiPhotonicMoleculesTailoring2014}. In this configuration, the coupling is predominantly reactive, leading to frequency splitting, while modal loss rates remain largely set by the individual cavities. As a result, dissipative coupling—governing loss redistribution between modes—cannot be tuned independently and typically requires unbalanced gain or absorption.
Moreover, direct coupling introduces cross-talk between resonators, hindering independent control of their frequencies and gain. 
An alternative approach consists in introducing a controlled phase in the coupling channel \cite{Peng2016,aslan2026coherent}, enabling interference between coupling pathways and access to both reactive and dissipative components. In particular, distant coupling mediated by a waveguide~\cite{madiot2024harnessing,razimanSingleModeEmissionPhaseDelayed2025,seegertSelfpulsingDynamicsMicroscopic2024} provides such phase control while spatially separating the resonators, thereby reducing cross-talk and enabling independent tuning of their parameters.

Here, we harness phase-controlled non-Hermitian coupling to implement active Q-switching in a pair of waveguide-coupled photonic crystal nanolasers. By maintaining one nanolaser in a continuous-wave subthreshold state and abruptly modulating the optical pump applied to the other, we induce a change in carrier density that modifies the detuning and interference conditions between the two cavities. This changes the Q-factor and net gain of the collective modes. When the resonance condition is crossed, one collective mode transiently enters a lower-loss, higher-gain state, leading to the rapid release of stored carrier energy as a short optical pulse.

In the following, we present a theoretical model describing the dynamics of two coupled nanolasers under asymmetric pumping, and identify the conditions under which resonant gain enhancement enables Q-switching (\cref{sec:theory}). We then experimentally demonstrate the generation of optical pulses in an indium phosphide photonic crystal platform and compare the measured results with numerical simulations (\cref{sec:exp}). Finally, we assess the bandwidth of this Q-switched regime, demonstrating over 6 GHz repetition rate (\cref{sec:BW}). 

\begin{figure}[!ht]
\centering
    \includegraphics[scale=1]{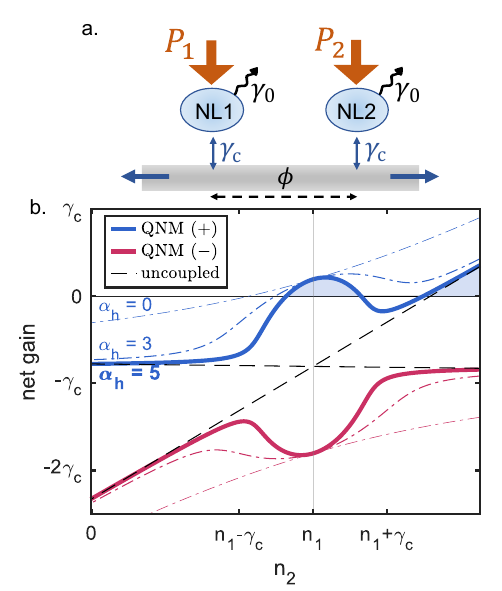}
    \caption{\textbf{a.} Schematic of the two phase-coupled nanolasers with identical decay rates and independent optical pump powers. \textbf{b.} Eigenmodes' ($+/-$) net gain parameter ($g_\pm$, in blue/red, respectively) as a function of the carrier density difference of NL2, $n_2$, while $n_1$ is kept constant to $0.65\times(1+\ntr)$. Black lines show the uncoupled cavity gain parameters. $g_\pm$ is computed using $\ah=5$ (thick line), and $\ah=3$ and $\ah=0$ (finer lines). Detuning $\delta$ is set to zero.}
    \label{Fig1}
\end{figure}

\section{Carrier-Mediated Non-Hermitian Coupling}
\label{sec:theory}
We consider two semiconductor laser nanocavities coupled to a unique optical waveguide, as depicted in \Cref{Fig1}.a. They are separated by a distance $L$ that produces a phase shift $\phi=2\pi L n_\mathrm{eff}/\lambda$, $\lambda$ being the wavelength and $n_\mathrm{eff}$ the effective waveguide index.  Writing $a_k$ and $n_k$ the complex amplitude of the electromagnetic field and the normalized carrier density in the cavity $k=\{1,2\}$, we establish the following coupled rate equations that describe the cavity amplitudes~\cite{manolatouCouplingModesAnalysis1999},

\begin{align}\label{eq:da_dt} 
    \dot{a}_{1,2} &= \frac{1}{2}\Big( -1 \pm i\delta+ (1+i\ah)(n_{1,2}-\ntr)\Big)a_{1,2} \\ 
    &- \gc e^{-i\phi}a_{2,1} + R_{1,2}^\mathrm{sp}(t) \nonumber
\end{align}

\begin{figure*}[!ht]
\centering
    \includegraphics[scale=1]{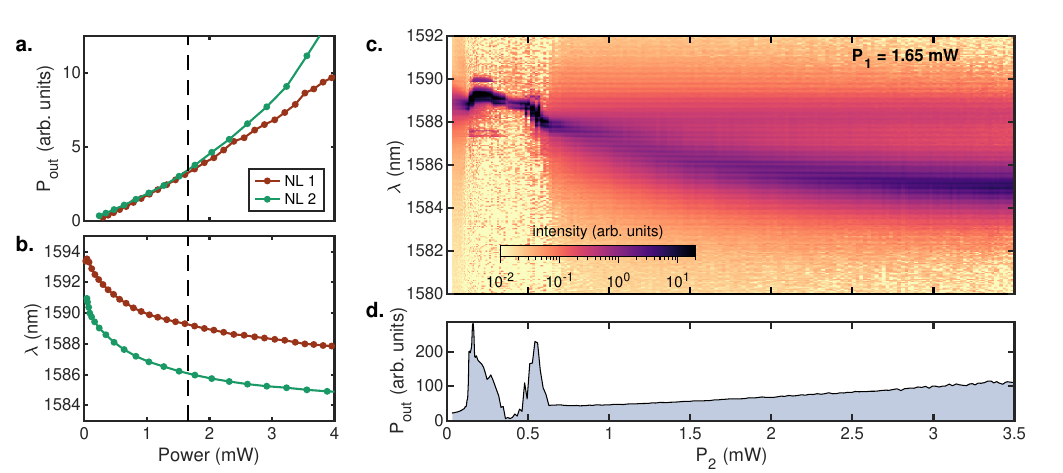}
    \caption{Experimental calibration of the independent nanolaser emission under optical pumping. \textbf{a.} The output powers and \textbf{b.} the associated emission wavelengths are shown as a function of the optical pump power applied to NL1 (red) or to NL2 (green). \textbf{c.} Measured output field spectrum as a function of $P_2$, while $P_1$ is fixed to $1.65$ mW. \textbf{d.} Output power as a function of P2, associated with the above measurement.}
    \label{Fig2}
\end{figure*}

where $\ntr$ is the carrier density at transparency and $\ah$ is the linewidth enhancement factor. The spontaneous emission rate, $R_{1,2}^\mathrm{sp}$, is neglected at this stage and will be expressed later. Details relative to the model normalization are given in \cref{ap:normalization}. In particular, time is rescaled to the photon lifetime $\tp=1/(2\Gamma_\mathrm{tot})$ with $\Gamma_\mathrm{tot}$ the total amplitude decay rate, which splits into internal and external cavity loss contributions, $\Gamma_\mathrm{0}$ and $\Gamma_\mathrm{c}$, respectively. Therefore, $\gc=\Gamma_\mathrm{c}\times\tp$ is the normalized coupling rate to the waveguide. We assume a detuning $\delta$ between the cavities, satisfying $\delta\ll \tp c/(Ln_\mathrm{eff})$ so that the chromatic dispersion of the phase shift can be neglected.

The case of $\phi= 0 \text{ or } \pi$ is of particular interest as it consists of anyonic PT-symmetry with Q-splitting between the electromagnetic QNMs \cite{madiot2024harnessing}. An eigenmode analysis for $\phi=0$ yields the net gain for each QNM:
\begin{align}
\label{eq:gain}
    g_\pm &= \frac{1}{2}(-1+\overline{n}-\ntr) \nonumber \\ 
    &\pm \Re\Bigg(\sqrt{\frac{1}{4}\Big(i\delta+(1+i\ah)\Delta n\Big)^2+\gc^2}\Bigg)
\end{align}

with the mean carrier density $\overline{n}=\frac{1}{2}(n_1+n_2)$ and the carrier density difference $\Delta n=\frac{1}{2}(n_1-n_2)$. We set $\delta=0$ to simplify the following discussion. This gain enhancement ($g_+$) or reduction ($g_-$) occurs under a resonance condition when both cavities are tuned, i.e., $\Delta n=0$.
Under this condition, the QNM ($+$) reaches a threshold carrier density of $n_\mathrm{th}=1+\ntr-2\gc$, matching the lasing threshold of an isolated nanolaser limited by internal losses ($\Gamma_0$).
This threshold reduction becomes significant when the external losses of the cavities dominate the internal losses. 
Crucially, resonant coupling can be tuned via the carrier density difference $\Delta n$, enabled by a non-zero Henry factor $\ah$. 

In \cref{Fig1}.b, we plot the QNM's gain, $g_\pm$, as a function of $n_2$, using $\phi=0$, $\gc=0.45$, and setting $n_1=0.65\times(1+\ntr)$ while $n_2$ is swept. The thick line corresponds to a realistic Henry factor, $\ah=5$. In this case, the QNM ($+$), in blue, is enhanced over the threshold (i.e. $g_+>0$) when the resonant condition on $\Delta n$ is fulfilled. Meanwhile, the other QNM ($-$), in red, is resonantly damped. We show the cases of $\ah=3$ and $\ah=0$ with thinner dashed lines, respectively showing weaker enhancement and no enhancement at all in the QNM gain increase or reduction. Finally, the two uncoupled cavity gains are shown with black lines. One is constant as it corresponds to the nanolaser 1 (NL1) whose carrier density, $n_1$, is kept unchanged, while the other corresponds to the nanolaser 2 (NL2) and grows linearly with $n_2$. For $\delta\neq0$, this behavior can be restored by unbalancing carrier densities, e.g., via asymmetric pump powers.

In the following, we exploit this resonant condition to dynamically cross the zero-detuning point by quickly modulating the carrier population in one nanolaser while the other is kept constant. Provided that both nanolasers are pumped below their respective lasing thresholds, stimulated emission will only be triggered when the resonant condition is fulfilled, i.e., when Q-switching occurs, leading to the generation of a light pulse. 

\section{Q-Switched Pulse Generation}
\label{sec:exp}
The experimental system consists of two indium phosphide photonic crystal nanocavities embedding four InGaAs quantum wells~\cite{bazinDesignSilicaEncapsulated2014,crosnierSubduingSurfaceRecombination2015}. Both are integrated into a silicon-on-insulator (SOI) waveguide, with their cavity centers separated by approximately 22 \micro m (see \cref{ap:device}). The external decay rate of each cavity to the SOI waveguide is $\Gamma_c\approx400$ GHz, while their internal decay is much lower, $\Gamma_0\approx 40$ GHz (see \cref{ap:transmission}). This results in very damped resonances, yielding poor lasing performance of the individual nanolasers. In \cref{Fig2}.a, we show the lasing curves of the independent nanolasers, that is, while the other is unpumped. The laser threshold is not visible below 5 mW for either NL1 or NL2, due to the high absorption losses. This is confirmed by the associated emission wavelengths, shown in \cref{Fig2}.b. For both nanolasers, as the pump power increases, the emission wavelength experiences a blue shift resulting from the free-carrier effect. Over the lasing threshold, the former would be compensated by a red shift due to the thermo-optical effect, which results from the increasing photon density in the cavity. The absence of such a red shift in the measured emission spectra indicates that the threshold has not been reached in this range of pump power. Due to the small cavity volumes, the carrier reservoirs are finite and, therefore, subjected to saturation, such that no lasing can be observed at any pump power with these nanolasers.

\begin{figure*}[!ht]
\centering
    \includegraphics[scale=0.99]{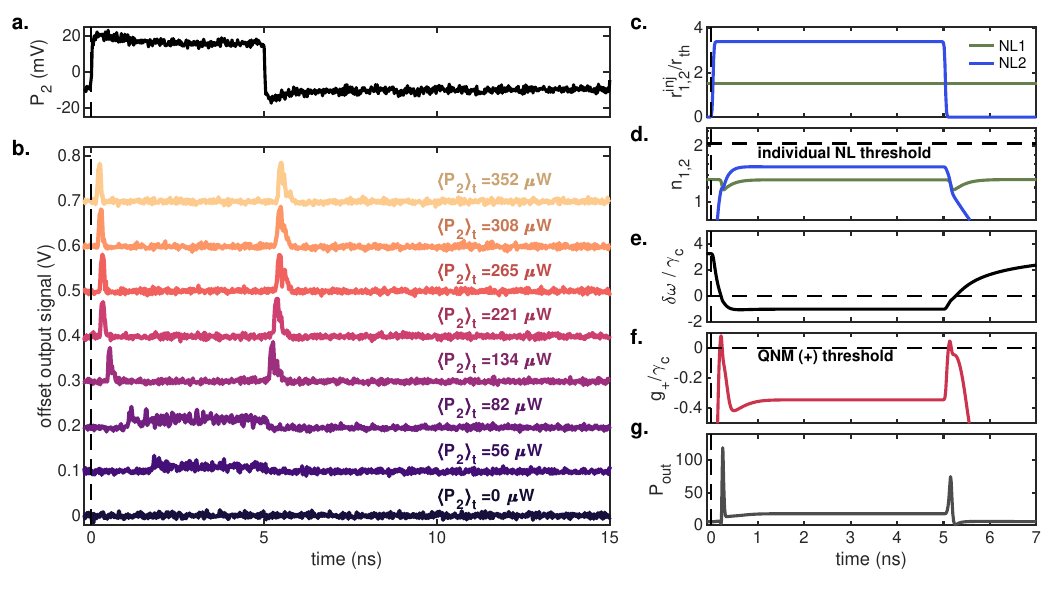}
    \caption{\textbf{a.} Time trace of the pump power $P_2(t)$, measured at the EOM output. \textbf{b.} Offset time traces of the output power, obtained with $P_1=3.04$ mW, and with increasing amplitude of the excitation pulse. The measured average pump power $\braket{P_2}_t$ is reported for each trace. Numerical simulation of the Q-switching process. \Cref{eq:da_dt,eq:dn_dt} are integrated and return \textbf{ c.}  the injection rates over time; \textbf{ d.} the carrier densities; \textbf{ e.} the frequency detuning; \textbf{ f.} the net gain of QNM ($+$), $g_+$; and \textbf{ g.}, the number of photons. The stationary state obtained when $\rinj_2=\rinj_1$ is shown with a red dashed line.}
    \label{Fig3}
\end{figure*}

Now we use both pumps and study the lasing emission of a collective mode. In \cref{Fig2}.c, we report the spectrum as a function of the pump power applied to NL2, $P_2$, while the pump power on NL1 is set to 1.65 mW.
A coherent and intense emission peak is produced within a short range of power (0.1-0.6 mW), when the two NLs are spectrally tuned. Higher pump power leads to a significant detuning between the nanolasers, which become uncoupled again. Two weaker and distinct emission peaks can be found: one that remains unchanged at 1588.41 nm as a function of $P_2$ and corresponds to NL1, while the other is slightly blue-shifted and is associated with NL2, similar to the independent measurement shown in \cref{Fig2}.b. Both peaks are also broader than in the intensity emission region, showing that they correspond to spontaneous emission. The total output power associated with this map is shown in \cref{Fig2}.d to illustrate the strong power enhancement when the collective emission is formed. The collective emission region includes a dip where the emission almost fall to zero. We attribute this extinction to destructive interference between the cavity leakages \cite{madiot2024harnessing}.

We demonstrate Q-switching by exploiting the above static phenomenon in a dynamical picture. For this, we continuously pump NL1 below its respective threshold, while NL2 is periodically pumped with $w=5$ ns long optical pump pulses. The nanolasers are independently pumped with a 1064 nm laser diode. The pump powers are finely controlled using electro-optical modulators (EOM). The pulse is generated in the RF domain using an arbitrary waveform generator and transduced into the optical pump via the corresponding EOM. The modulated pump's rise and fall times (\cref{Fig3}.a) are below the 30 ps resolution limit of the oscilloscope. 
The emission of the nanolasers is collected by aligning an optical fiber to a waveguide termination, formed by grating couplers. The collected light is amplified in an erbium-doped fiber amplifier (EDFA) and filtered with a Fabry-Pérot tunable bandpass filter. The optical signal is then sent to a 40 GHz bandwidth InGaAs photodetector and read by the oscilloscope.

In \cref{Fig3}.b, we show the measured output signal obtained for increasing values of the average pump pulse power $\braket{P_2}_t$, while the CW pump power is constant and fixed to $P_1=3.04$ mW. The data are plotted accounting for a calibrated time-delay in the setup (see \cref{ap:delay}). We observe a subthreshold regime, when $\braket{P_2}_t<56$ \micro W, where the signal is weak and constant over time. It cannot be distinguished from a situation where the pump lasers are turned off. For $\braket{P_2}_t \sim 56$-$82$ \micro W, the pump pulse has an amplitude such that the nanolasers are nearly tuned during the 5 ns pulse. This results in nanolaser coupling and a net gain for the QNM $(+)$ , which converges towards a stationary lasing state. Oscillatory regimes and instabilities are observed within this range, due to proximity to the QNM threshold. 
For higher values of $P_2$, two optical pulses are observed when the excitation pulse rises and falls, respectively. The latency between the excitation pulse ($t=0$) and the response pulse tends to reduce for increasing values of $P_2$, as discussed in the following. Meanwhile, the emission pulse width varies from 70 ps to 300 ps.

To understand the mechanism of output pulse formation, we perform numerical simulations of the coupled nanolasers system. As the nanolasers follow a class-B semiconductor laser model, the carrier densities are described by the following rate equations:

\begin{align}\label{eq:dn_dt}
    \dot{n}_{1,2} &= \rinj_{1,2}(t)\Big(1-\frac{n_{1,2}}{\nsat}\Big) - \frac{n_{1,2}}{\tnr} - \frac{n_{1,2}^2}{\trad} \\& - g_0(n_{1,2}-\ntr)|a_{1,2}|^2\nonumber
\end{align}

where $\tnr$ and $\trad$ are the normalized nonradiative and radiative carrier decay times, respectively. $g_0$ is the differential gain and $\rinj_{1,2}$ are the carrier injection rates. In addition, we introduce a saturation of the carrier densities through the saturation density $\nsat$, in agreement with the observations in \cref{Fig2}a-b. \Cref{eq:da_dt} is computed using a deterministic spontaneous emission rate $R_{1,2}^\mathrm{sp}(t) = \eta n_{1,2}^2 \frac{a_{1,2}}{|a_{1,2}|^2}$, where $\eta$ depends on the nanolasers' properties and is set to $\eta=2.34$. The intrinsic frequency detuning is obtained from the CW characterization reported in \cref{Fig1}, yielding $\delta=-0.347$.

\begin{figure}[!ht]
\centering
    \includegraphics[scale=1]{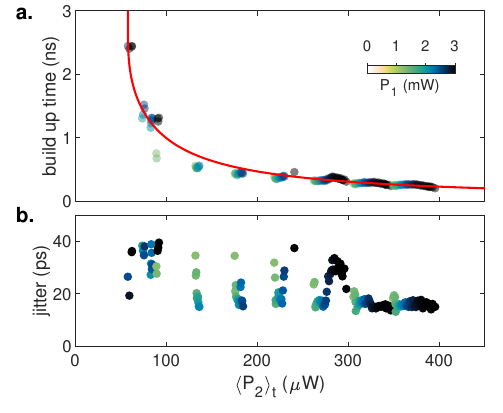}
    \caption{\textbf{a.} Upward pulse arrival time as a function of $P_2$, and theoretical fit (red line). \textbf{b.} Associated timing jitter computed for each trace as the standard deviation of the arrival times. The measurements are obtained with different pump power $P_1$ (refer to colorbar).}
    \label{Fig4}
\end{figure}

We resolve \cref{eq:da_dt,eq:dn_dt} using $\rinj_1=1.5\times r_\mathrm{th}$, and $\rinj_2(t)=3.5\times r_\mathrm{th}\times f(t)$, where $f(t)$ is a pulse function with exponential rise and fall decays, and such that the pulse starts at $t=0$. $r_\mathrm{th}$ is the normalized injection rate threshold of an individual nanolaser in the absence of saturation ($\nsat\gg n_\mathrm{th}$). 
In \cref{Fig3}.c, the injection rates of nanolasers 1 and 2 are represented in green and blue, respectively. The associated carrier densities (d) show latency in their trajectory compared to the pump due to long carrier recombination timescales $\tnr\times \tp=3.2$ ns and $\trad\times \tp=1.55$ ns.
The instantaneous frequency detuning (e), given by $\delta\omega=\delta+\ah\Delta n/2$, crosses zero around t=800 ps, leading to an enhancement of $g_+$ (f). This corresponds to the Q-switching operation, and the carriers are quickly depleted into photons, leading to a short and intense optical pulse in the waveguide output power (g). This emission process occurs on a timescale given by the photon decay rate, which is much faster than the carrier dynamics. 
The same process occurs during the extinction of the excitation pulse. Importantly, the initial conditions are now different and allow for a higher load of carriers in NL2, compared to the upward scenario. This results in a slightly stronger optical downward pulse.

The upward pulse build-up time is defined as the time between the excitation pulse absorption by the NL2 and the Q-switched pulse start. For a given set of pump powers, a time series is recorded, so that the build-up time can be averaged over several events, and the associated standard deviation provides the timing jitter of the Q-switched pulses.
The build-up time is plotted in ~\cref{Fig4}.a as a function of the average power $\langle P_2\rangle_t$. The measurements are obtained at different values of $P_1$, as indicated by the colorbar. We observe no significant dependency on $P_1$. The build-up time shows a stiff decay from 2.5 ns at $\langle P_2\rangle_t\approx60$ $\mu$W, down to $0.2$ ns where it saturates for $\langle P_2\rangle_t>300$ $\mu$W. The data is fitted with an analytic function obtained from \cref{eq:dn_dt}, using $\nsat$ and a linear absorption coefficient of the pump power by NL2 as fitting parameters, as described in \cref{ap:fit}. 
In ~\cref{Fig4}.b, we report the associated timing jitter, which ranges from 15 to 40 ps and tends to decrease with increasing pump power $P_1$.

\begin{figure}[!ht]
    \centering
    \includegraphics{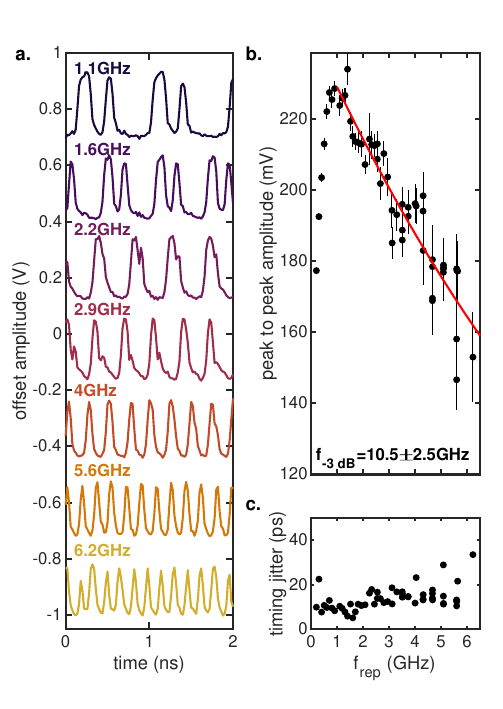}
    \caption{\textbf{a.} Measured output signal under square modulation input, for varying repetition rate $T_\mathrm{rep}=f_\mathrm{rep}^{-1}$ and using a duty cycle of 50\%.  Over each trace are computed \textbf{b.} the peak-to-peak amplitude with associated amplitude jitter, and \textbf{c.} the timing jitter.}
    \label{fig:5}
\end{figure}

\section{High-Frequency Operation}
\label{sec:BW}
To evaluate the performance of the Q-switched pulses at higher repetition rates, we drive NL2 with a square modulation signal at 50\% duty cycle and record the temporal response for increasing repetition frequency $\frep$. Representative time traces are shown in Fig.~5(a). All measurements are performed using identical amplification, filtering, and detection settings.

For $f_{\mathrm{rep}} \lesssim 2.7~\mathrm{GHz}$, each modulation period produces two distinct pulses, corresponding to the upward and downward Q-switching transitions. As the repetition rate increases, the system transitions to a single-pulse-per-period regime. This shift arises because the carrier population fails to fully recover between successive pulses, suppressing the second Q-switching event, as discussed in \cref{ap:sim}.
For each repetition rate, we extract the pulse amplitudes and arrival times over multiple periods. The mean peak-to-peak amplitude is plotted as a function of $f_{\mathrm{rep}}$ in \cref{fig:5}.b, with error bars corresponding to the amplitude jitter given by the associated standard deviation. Below $\frep=1$ GHz, two Q-switched pulses are observed for each period, with increasing peak-to-peak amplitudes. For higher repetition rates, the peak-to-peak amplitude decreases. We fit an exponential decay curve to the data above $1$ GHz to estimate the cut-frequency at which the signal amplitude reduces by 3 dB (red curve). We find $f_\mathrm{-3 dB}=10.5 \pm 2.5$ GHz.

The average temporal separation between successive pulses yields the modulation period $1/\frep$, while the standard deviation of this delay provides the timing jitter, shown in \cref{fig:5}c. The timing jitter fluctuates between $5$ and $40~\mathrm{ps}$ across the investigated frequency range, in agreement with the values obtained in the single-pulse measurements (see \cref{Fig4}.b). 

Overall, this experiment shows that Q-switched operation is maintained up to multi-gigahertz modulation frequencies, with a clear transition from double-pulse to single-pulse emission governed by carrier recovery dynamics. The extracted cut-off frequency indicates a substantial effective bandwidth for this collective nanolaser system. Despite the reduction in pulse amplitude at high repetition rates, the timing jitter remains limited, demonstrating stable temporal operation over the entire investigated range.

\section{Conclusion}
We have demonstrated active Q-switching in a pair of phase-coupled nanolasers by exploiting non-Hermitian gain modulation through asymmetric optical pumping. By dynamically tuning the carrier-induced detuning between the cavities, we transiently activate a collective mode with enhanced net gain, enabling the emission of short optical pulses from nanocavities that individually do not sustain efficient continuous-wave lasing. The pulse build-up dynamics, energy, and timing jitter are well captured by a class-B rate-equation model including carrier saturation, confirming that the operation is governed by carrier recovery rather than photon lifetime. Repetition-rate measurements reveal stable operation up to multi-gigahertz frequencies, with an effective modulation bandwidth exceeding $10~\mathrm{GHz}$. 

These results establish non-Hermitian coupling as a powerful mechanism for engineering ultrafast pulse generation in integrated nanophotonic platforms, opening new perspectives for compact on-chip sources in high-speed communication, signal processing, and laser network architectures.

\begin{acknowledgments}
This work was partly supported by the European Research Council (ERC) project HYPNOTIC (grant agreement number 726420), by France 2030 government grants (ANR-22-PEEL-0010 and ANR-15-IDEX-01), by the French National Research Agency (ANR), Grants No ANR-25-CE24-1834 and ANR-23-CE24-0013, and by the RENATECH network. K.S. acknowledges the support from the Danish National Research Foundation (Grant No. DNRF147 - NanoPhoton) and from the European Research Council (Grant No. 834410 FANO). 
\end{acknowledgments}
\section*{Data availability statement}
The data that support the findings of this article are openly available~\cite{seegertDynamicalControlNonHermitian2026}.
\crefname{section}{appendix}{appendices}  
\appendix

\section{Normalization of the rate-equations}
\label{ap:normalization}
We consider two nanolasers that interact distantly, via a waveguide. The rate equations describing the complex amplitude and associated carrier density in cavity $1,2$ read

\begin{align}
\dot{\alpha}_{1,2} &= \Big( i\omega_{1,2} - (\Gamma_c+\Gamma_0) \nonumber + (1+i\ah)\frac{V_aG(N_{1,2})}{2}\Big) \alpha_{1,2} \nonumber\\ &- \Gamma_c e^{-i\phi} \alpha_{2,1}  + \frac{\beta V_aF_pBN_{1,2}^2}{2}\frac{\alpha_{1,2}}{|\alpha_{1,2}|^2} \\
\dot{N}_{1,2} &= \Rin_{1,2} - F_p B N_{1,2}^2 - \frac{N_{1,2}}{\uptau_\mathrm{nr}}-G(N_{1,2})|\alpha_{1,2}|^2
\end{align}

with $V_a$ the active volume, $\beta$ the spontaneous emission rate, $F_p$ the Purcell enhancement factor, $B$ the bimolecular recombination rate, and $\uptau_\mathrm{nr}$ the non-radiative carrier lifetime. $\Rin_{1,2}$ is the injection rate in cavity $1,2$.
We rewrite the above equation in a frame rotation at the average of the cavities' frequencies, $\overline{\omega}=\frac{1}{2}(\omega_1+\omega_2)$. The nanolaser gain is linearized as $G(N)\approx G_0(N/N_\mathrm{tr}-1)$, and with $G_0=\beta F_p B/A'$ is the gain factor that depends on a material's constant $A'$.

Time is normalized by the photon lifetime $\tp=\frac{1}{2}(\Gamma_0+\Gamma_c)^{-1}$. Carrier densities are also renormalized using $n_{1,2}= N_{1,2}/N_\mathrm{x}$ with $N_\mathrm{x}=N_\mathrm{th}-N_\mathrm{tr}$ i.e. the amount of carrier density required to go from transparency to threshold can be evaluated as $N_\mathrm{x}=N_\mathrm{tr}/(V_aG_0\tp)$. Finally, in order to establish \cref{eq:da_dt,eq:dn_dt}, we introduce the following normalized parameters:
$\tnr=\uptau_\mathrm{nr}/\tp$
 $\gc =\Gamma_c \tp$, 
$\ntr=N_\mathrm{tr}/N_\mathrm{x}=V_aG_0\tp$, 
$\rinj_{1,2}= \tp \Rin_{1,2}/N_\mathrm{x}$, 
$g_0=\tp G_0/N_\mathrm{tr}$, 
$\trad=(N_\mathrm{x} F_p B \tp)^{-1}$, and
$\eta=(\tp\beta V_a F_p B N_\mathrm{x}^2)/2$.

The threshold carrier density of an individual cavity without carrier saturation ($\nsat\gg n_\mathrm{th}$) is $N_\mathrm{th} = N_\mathrm{tr}(1+n_x)$, by definition. We deduce the normalized injection rate at threshold $r_\mathrm{th}=(\ntr+1)/\tnr+(\ntr+1)^2/\trad$. 
 The set of physical parameters used in the simulations is shown in \cref{tab:phys_quant}.

\begin{table}[!h]
    \centering
    \begin{tabular}{c|c|c}
        symbol & meaning & value \\
 \hline 
$\Gamma_0$ & internal decay rate    & 40 GHz    \\
$\Gamma_c$ & external decay rate    & 400 GHz    \\
$\tp$  & photon lifetime        & 1.14 ps  \\
$\beta$ & spontaneous emission constant & 0.1  \\
$\uptau_\mathrm{nr}$ & nonradiative carrier lifetime & 2.3 ns \\
$N_\mathrm{tr}$ & carrier density at transparency & $10^{18}$ cm$^{-3}$ \\
$\ah$ & Henry factor & 5 \\
$F_p$ & Purcell enhancement factor & 2.3 \\
$B$ & bimolecular recombination rate & $3\times10^{-10}$ cm$^{3}$/s \\
$A'$ & material constant & 5$\times 10^{-36}$ cm$^6$ cm$^{-3}$ \\
$V_a$ & cavity active volume & 6.8$\times10^{-14}$ cm$^3$ \\
$G_0$ & Gain factor, $G_0=\beta F_p B/A'$ & 1.38$\times10^{25}$ cm$^{-3}$/s \\
    \end{tabular}
    \caption{Set of physical parameters used in the numerical simulations}
    \label{tab:phys_quant}
\end{table}

\section{III-V on SOI integrated nanolaser array}
\label{ap:device}
\begin{figure}
    \centering
    \includegraphics{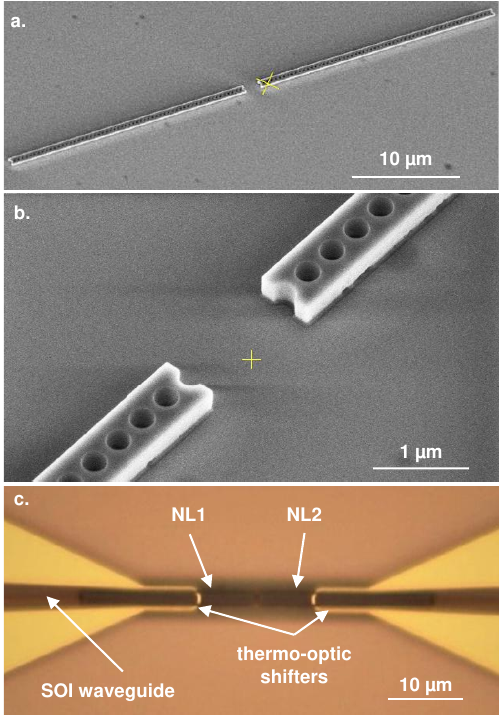}
    \caption{\textbf{a.} SEM image of the array and \textbf{b.}, zoom-in on the photonic crystal cavities inward terminations. \textbf{c.} Optical microscope image of the final system: III-V nanolasers (NL1 and NL2) integrated on a SOI waveguide, underneath gold nanowires for thermo-optic control.}
    \label{fig:structures}
\end{figure}

Scanning electron microscope (SEM) images of the sample are provided in \cref{fig:structures}, highlighting the full array of two coupled nanolasers (a) and a zoom-in of the photonic-crystal nanocavity edges, with a pitch distance of approximately 1.3 \micro m and a center-to-center distance of approximately 22 \micro m. These images are obtained after plasma etching, prior to encapsulation of the nanostructures with silica. A final step consists in the integration of gold nanowires on top of each nanolaser, enabling thermo-optic control over their resonances. An optical microscope image of the final system is shown in \cref{fig:structures}.c.

\section{Waveguide-coupling rate calibration}
\label{ap:transmission}
For a single nanolaser onto a SOI waveguide, the normalized transmission spectrum is given by:
\begin{equation} \label{eq:Tra}
    T(\Delta) \equiv \frac{P_\mathrm{out}}{P_\mathrm{in}}= \frac{\Delta^2+(\Gamma_\mathrm{tot}-\Gamma_c)^2}{\Delta^2+\Gamma_\mathrm{tot}^2}
\end{equation}

where $\Delta=\nu_L-\nu_0$ is the frequency detuning between the laser ($\nu_L$) and the cavity resonance ($\nu_0$), and $\Gamma_\mathrm{tot} = \Gamma_0+\Gamma_\mathrm{abs}+\Gamma_c$ is the total amplitude decay rate of the cavity. It is composed of the cavity internal decay rate at transparency ($\Gamma_0$), the cavity loss rate resulting from absorption in the quantum wells ($\Gamma_\mathrm{abs}$), and the external decay rate to the integrated waveguide ($\Gamma_c$). Fitting the measured normalized transmission with the above expression can therefore provide unambiguous characterization of $\Gamma_c$.

\begin{figure}[!ht]
    \centering
    \includegraphics{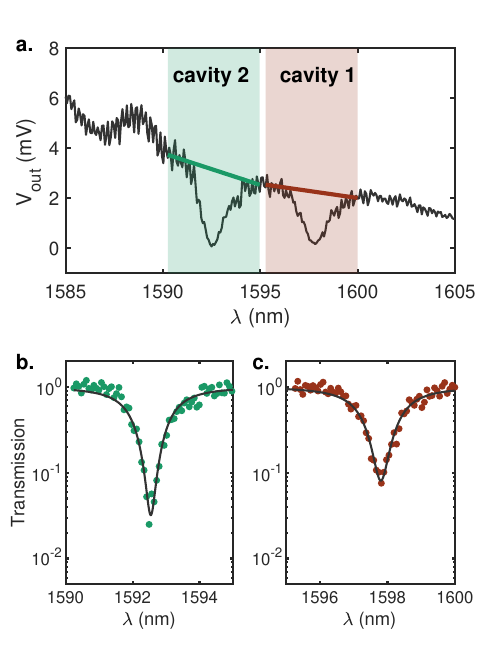}
    \caption{\textbf{Calibration of the waveguide coupling rates}. a. We measure the waveguide transmission with a tunable laser. Two dips are identified and attributed to the nanolaser cavities' resonance. The off-resonance noise signals levels are fitted with linear functions (colored curves) to obtain the normalized transmission. b-c. Both transmission dips are fitted (black curves), providing the values of $\Gamma_c$ for each cavity.}
    \label{fig:gamma_c}
\end{figure}

We measure the transmission spectrum of the waveguide using a tunable laser at telecommunication wavelengths. The laser is polarized using a fiber polarization controller to maximize the transverse-electric cavity mode energy. This is verified by tuning the laser to resonance and maximizing the intensity of the cavity scattering with a microscope coupled to an infrared camera. Then, the laser wavelength is scanned over a large span covering both nanolaser resonances. A photodetector converts the output signal into a voltage, which is visualized on an oscilloscope. Time is converted into wavelength via the laser scan speed, which is set to 5 nm/s. In order to avoid mode-coupling between the cavities, the nanolaser 1, which is already red-detuned as shown in \cref{Fig2}.b, is thermo-optically red-shifted by few nanometers. This is achieved through integrated gold nanowires \cite{madiot2024harnessing}. The resulting output spectrum is shown in \cref{fig:gamma_c}.a).
Focusing independently on both resonance dips (shaded blue and green regions) , we normalize the output spectrum by a linear background evaluated off-resonance (blue and green lines). 
The normalized transmission spectra (\cref{fig:gamma_c}.b-c) are fitted with \cref{eq:Tra} (red lines), returning the cavities' external coupling rates $\Gamma_c^\text{cavity 1} = 440\pm28$ GHz and $\Gamma_c^\text{cavity 2}=396\pm22$ GHz, where the uncertainty corresponds to the fit 95\% confidence interval. In the model and subsequent numerical simulations, we assume both external decay rates to be equal, and set to 400 GHz.

\section{Pulse delay calibration}
\label{ap:delay}

\begin{figure}[!ht]
    \centering
    \includegraphics[width=1\linewidth]{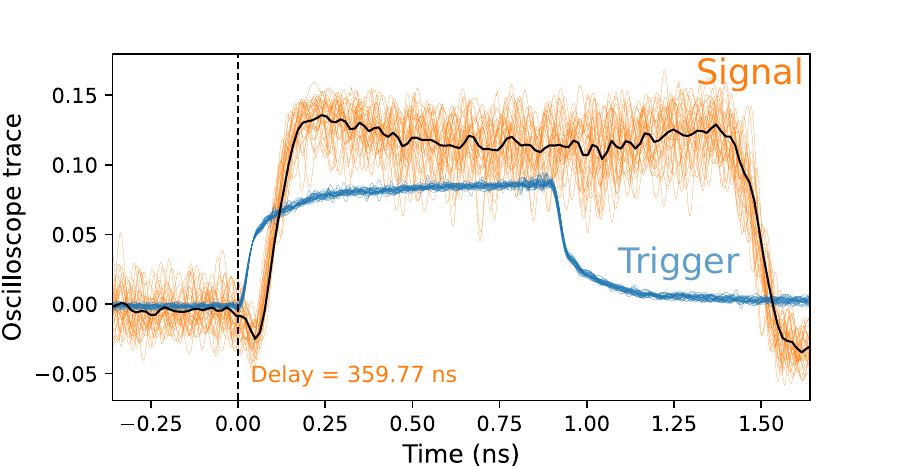}
    \caption{Calibration of the zero-time delay, showing the superposed signal (orange) and trigger (blue) oscilloscope traces. The black curve shows the averaged signal trace.}
    \label{fig:delay_calib}
\end{figure}

The RF pulses are generated by the arbitrary waveform generator (AWG). A second AWG output is used to trigger the oscilloscope. On the detection side, the emitted pulses are amplified in an EDFA, which introduces a significant delay in the oscilloscope trace between the arrival of the trigger pulse and the corresponding emitted pulse.
To obtain the build-up time, we calibrate the relative arrival time of the pump pulse at the sample. We do this by injecting a tunable laser source at one end of the waveguide and measuring the transmission at the other. The zero-time delay is then defined as the delay in the oscilloscope trace between the onset of the trigger pulse and the earliest change in the signal response. We calibrate this to $t_0=359.774(\pm0.015)$ ns. Figure \ref{fig:delay_calib} shows the superposed trigger and signal traces used in the calibration. 

\section{Pulse build-up time}\label{ap:fit}

We compute the time $\Delta t$ necessary for the emission of a Q-switched pulse. It is given by integrating the carrier master equation in \cref{eq:dn_dt} between an initial state $n_\mathrm{i}$ and a final state $n_\mathrm{f}$, and assuming no stimulated emission ($g_0=0$).
\begin{align}
\label{Deltat}
    \Delta t &=  \tp\times\int_{n_\mathrm{i}}^{n_\mathrm{f}} \frac{\mathrm{d}n_2}{\rinj_2-n_2/\teff - n_2^2/\trad} \nonumber\\ 
   &= \frac{\tp}{\gamma_2}\,
\ln\!\left[
\frac{\bigl(n_\mathrm{f}-n_{2,+}\bigr)\bigl(n_\mathrm{i}-n_{2,-}\bigr)}
     {\bigl(n_\mathrm{f}-n_{2,-}\bigr)\bigl(n_\mathrm{i}-n_{2,+}\bigr)}
\right]. 
\end{align}

with $\gamma_2 \equiv \sqrt{1/\teff^{2}+4\rinj_2/\trad}$, and 
$n_{2,\pm}\equiv \frac{1}{2}\trad\left(-1/\teff\pm \gamma_2\right)$. We have introduced an injection-rate dependent carrier lifetime, which accounts for the carrier density saturation, $\teff^{-1}=\tnr^{-1}+\rinj_2/\nsat$.
We take $n_\mathrm{i}=0$, that is, considering that there is no carrier in NL2 when the associated pump is off. The pulse starts when $n_2\sim n_1$, which can be evaluated in the stationary state of NL1, $n_\mathrm{f}=n_1\approx \rinj_1 \tnr/\trad$. We relate the pump power with the injection rate through a linear coefficient $\coeff=P_2/\rinj_2$. We scale $\Delta t$ to the photon lifetime $\tp=1.03$ ps and fit it to the data in \cref{Fig3} using $\coeff$ and $\nsat$ as fitting parameters. In particular, we find $\nsat=1.9$, which lies below the individual threshold carrier density ($n_\mathrm{th}=N_\mathrm{tr}+1\approx2.07$). This indicates that the nanolasers can not reach their individual threshold, and is consistent with the observation of the lasers' response under CW pumping reported in \cref{Fig1}. This value is used in the numerical simulations reported in \cref{Fig3}.c-g.

\section{Two-pulses and one-pulse regimes in simulations}\label{ap:sim}

\begin{figure}[!ht]
	\centering
	\includegraphics[width=8.5cm]{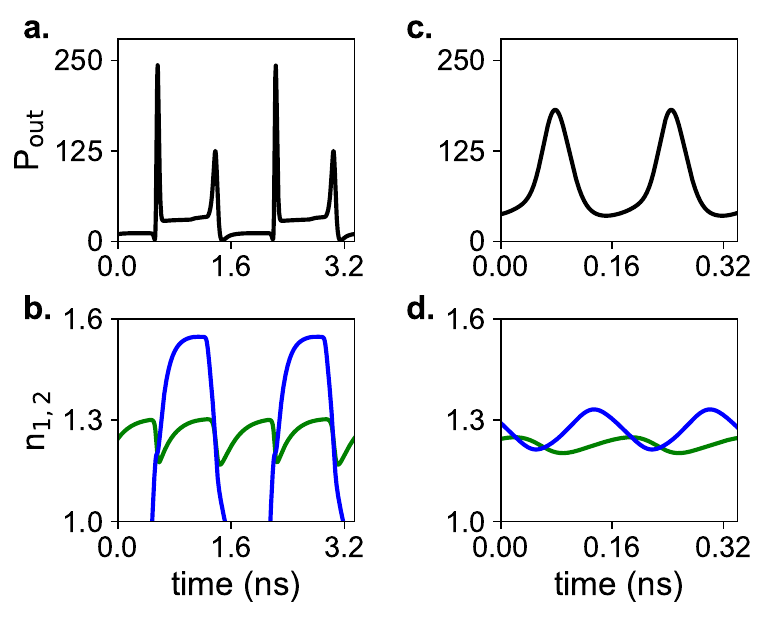}
	\caption{Numerical simulation of the Q-switching process. \Cref{eq:da_dt,eq:dn_dt} are integrated using a modulation pulse with f$_{\text{rep}}=0.6$ GHz (left) and f$_{\text{rep}}=6$ GHz (right). \textbf{a.} and \textbf{c.} show the total number of photons; and \textbf{b.} and \textbf{d.} show the carrier densities.}
	\label{fig:two_pulses-one_pulse}
\end{figure}

To understand the carrier density behavior in the two-pulse and one-pulse regimes (\cref{fig:5}.a), we simulated \cref{eq:da_dt,eq:dn_dt} over a range of repetition frequencies, f$_{\text{rep}}$, of the modulation input, using a duty-cycle of 50\%. The results are summarized in \cref{fig:two_pulses-one_pulse}. As previously discussed, the Q-switching mechanism relies on switching from non-resonant ($\Delta n\neq0$) to resonant ($\Delta n=0$) coupling between the individual cavities. At low repetition frequencies (\cref{fig:two_pulses-one_pulse}a,b) the period of the modulation pulse is long enough to allow the carrier densities $n_{1}$ and $n_{2}$ to reach a non-resonant state twice per period: once when $n_{2}$ saturates above $n_{1}$ (modulation on) and again when $n_{2}$ relaxes below $n_{1}$ (modulation off). This results in two generated pulses per cycle with a full width at half maximum (FWHM) of $0.040$~ns. However, at high repetition frequencies (\cref{fig:two_pulses-one_pulse}c,d) the modulation period is too short for $n_{2}$ to saturate; consequently, one pulse disappears, leaving only the pulse associated with $n_{2}>n_{1}$, which exhibits a FWHM of $0.046$~ns. In particular, \cref{fig:two_pulses-one_pulse}d shows that the single pulse generated in the high-frequency regime is created within the time window when $\Delta n\approx0$.

\newpage

\bibliography{refs}

\end{document}